\documentclass[graybox, nosecnum]{svmult}

\usepackage{mathptmx}       
\usepackage{helvet}         
\usepackage{courier}        
\usepackage{type1cm}        
\usepackage{mathtools,amsfonts,amssymb}
\usepackage{makeidx}         
\usepackage{graphicx}        
\usepackage{multicol}        
\usepackage[bottom]{footmisc}
\usepackage{hyperref}        
\usepackage{soul}            

\makeindex             

\usepackage[numbers]{natbib}
\usepackage{etoolbox}
\usepackage{hyperref}

\makeatletter
\newlength{\longestbiblabel}
\newlength{\runninglongestbiblabel}

\newcommand*{\abx@longestbiblabel}[1]{%
  \global\setlength{\longestbiblabel}{#1}}

\patchcmd\NAT@bibsetnum
  {\setlength{\leftmargin}{\labelwidth}}
  {\setlength{\labelwidth}{\longestbiblabel}%
   \setlength{\leftmargin}{\labelwidth}}
  {}{}

\newcommand\pyser@setlabwidth[2]{%
  \begingroup
  \settowidth{\@tempdima}{\@biblabel{#2}}%
  \ifnum\@tempdima>#1%
    \global#1\@tempdima
  \fi
  \endgroup}

\renewcommand\NAT@wrout[5]{%
  \pyser@setlabwidth{\runninglongestbiblabel}{#1}%
  \if@filesw
    {\let\protect\noexpand\let~\relax
     \immediate\write\@auxout{%
       \string\bibcite{#5}{{#1}{#2}{{#3}}{{#4}}}}}%
  \fi
  \ignorespaces}

\AtEndDocument{%
  \if@filesw
     {\let\protect\noexpand\let~\relax
      \immediate
      \write\@auxout{%
        \string\abx@longestbiblabel{\the\runninglongestbiblabel}}}%
  \fi
  \ifdim\longestbiblabel=\runninglongestbiblabel
  \else
    \G@refundefinedtrue
  \fi
}
\makeatother

\begin{document}
\title*{Modern approaches to optical potentials}
\author{Jeremy W.\ Holt\thanks{corresponding author} and Taylor R.\ Whitehead}
\institute{Jeremy W.\ Holt \at Department of Physics and Astronomy and Cyclotron Institute, Texas A\&M University, College Station, TX 77843, \email{holt@physics.tamu.edu}
\and Taylor R.\ Whitehead \at Facility for Rare Isotope Beams, Michigan State University, East Lansing, MI 48824, \email{whitehead@frib.msu.edu}}
%
%
\maketitle
\abstract{
This chapter presents an overview of the optical model description of nucleon-nucleus scattering and reactions based on fundamental nuclear two-body and many-body forces. The chapter begins with a historical review followed by a discussion of several commonly used theoretical techniques for deriving nucleon-nucleus optical model potentials that account for antisymmetry, Pauli blocking, and multiple scattering. This is followed by a summary of current efforts to derive microscopic nuclear forces consistent with the fundamental theory of strong interactions, quantum chromodynamics, and the use of such interactions in the construction of microscopic optical potentials. The chapter will also outline the dispersive optical potential approach, which despite being primarily phenomenological, has its origins in formal Green's function theory. Finally, the results from modern optical potentials will be benchmarked to experimental data.
}

\section{\textit{Introduction}}
Nuclear reactions are fundamental for understanding the chemical evolution of the universe as well as energy generation in stars. On Earth, nuclear reactions find applications to national security, medical diagnostics and therapy, and nuclear power. Nuclear reaction experiments are also used in basic science research to probe the structure of atomic nuclei and the properties of the strong nuclear force. In the coming decades, a major research direction in the field of nuclear science will be to explore unknown regions of the nuclear chart through next-generation rare-isotope beam facilities. The masses, decays, and reactions of these exotic isotopes will help elucidate subtle effects from proton-neutron asymmetries in nuclei and nuclear matter with implications for neutron star physics as well as novel nucleosynthesis in transient stellar phenomena such as core-collapse supernovae and neutron star mergers.

The interpretation of reaction experiments involving exotic isotopes will largely rely on theoretical {\it models}, rather than on ab initio many-body calculations, since in most cases the exact solution of the quantum many-body problem for a projectile nucleon incident on a target nucleus is intractable. This is complicated by the numerous exit channels that may occur in a nucleon-nucleus scattering experiment:
\begin{align}
A + N & \longrightarrow A + N & \text{Elastic scattering} \\
A + N & \longrightarrow A^* + N^* & \text{Inelastic scattering} \\
A + p & \longrightarrow A^\prime + n & \text{Charge exchange} \\
A + N & \longrightarrow (A + N) + \gamma & \text{Radiative capture}\\
(B+x) + N & \longrightarrow B + (N + x) & \text{Transfer} \\
(B+x) + N & \longrightarrow B + N + x & \text{Breakup}
\end{align}
where $A$ denotes an arbitrary nucleus, $N$ denotes an arbitrary nucleon, $p/n$ denotes a proton/neutron, a star superscript denotes a change in energy, and elements in parentheses denote the components of a nucleus.

The {\it optical model} is just one approach to analyze nucleon elastic scattering and reaction cross sections. The virtue of the model is its simplicity: although the nuclear force is characterized by complicated two-body, three-body, four-body, etc.\ components, the interaction of a projectile nucleon with a target nucleus can be surprisingly well approximated by potential scattering from a complex, nonlocal, and energy-dependent {\it one-body potential}. This one-body potential arises by averaging over the more fundamental nucleon-nucleon and many-nucleon components of the nuclear force and by projecting out the inelastic scattering channels. The optical model is therefore the positive-energy analogue of the nuclear shell model, which similarly describes the ground and excited (negative energy) states of a nucleus in terms of an average one-body mean field potential with associated single-particle orbitals filled by the nucleons. Despite its simplicity, the optical model also encodes complicated quantum many-body effects including antisymmetry, Pauli blocking, virtual excitations and correlations. However, the model does not resolve the fluctuations in scattering cross sections due to individual states or resonances in the combined nucleus, and therefore at low projectile energies the model can only provide a description of energy-averaged (or low-resolution experiment) cross sections. The optical model can also be extended to describe inelastic nucleon-nucleus reactions by singling out specific exit channels that are treated together with elastic scattering within a coupled channels formalism \cite{tamura65}.

The rest of this chapter will focus on recent developments in the construction of nucleon-nucleus optical model potentials. Starting from a brief overview of the phenomenological approach that has undergone continuous refinement over many decades, the discussion turns to the different formal methods have been developed to derive the effective interaction describing nucleon-nucleus elastic scattering starting from quantum many-body theory incorporating microscopic nuclear forces. These include Feshbach's projection formalism \cite{feshbach58}, Watson's multiple scattering theory \cite{watson53} and its later extensions \cite{kerman59}, and the Green's function approach \cite{bell59}. The motivation behind these modern approaches based on microscopic many-body theory is to enable a more consistent description of nuclear structure and reactions connected by fundamental theories of the nuclear force. The dispersive optical model \cite{mahaux86} will also be discussed. This formalism is similarly motivated by the need for a more consistent framework for understanding nuclear scattering and nuclear structure data, though from a primarily phenomenological perspective. Given the long history of the topic, there are numerous excellent review articles \cite{feshbach58,hodgson71,jeukenne76,mahaux91,dickhoff18} devoted to the nuclear optical model and which can be referred to for additional details and expanded discussions.

\begin{figure}[t]
\includegraphics[scale=0.74]{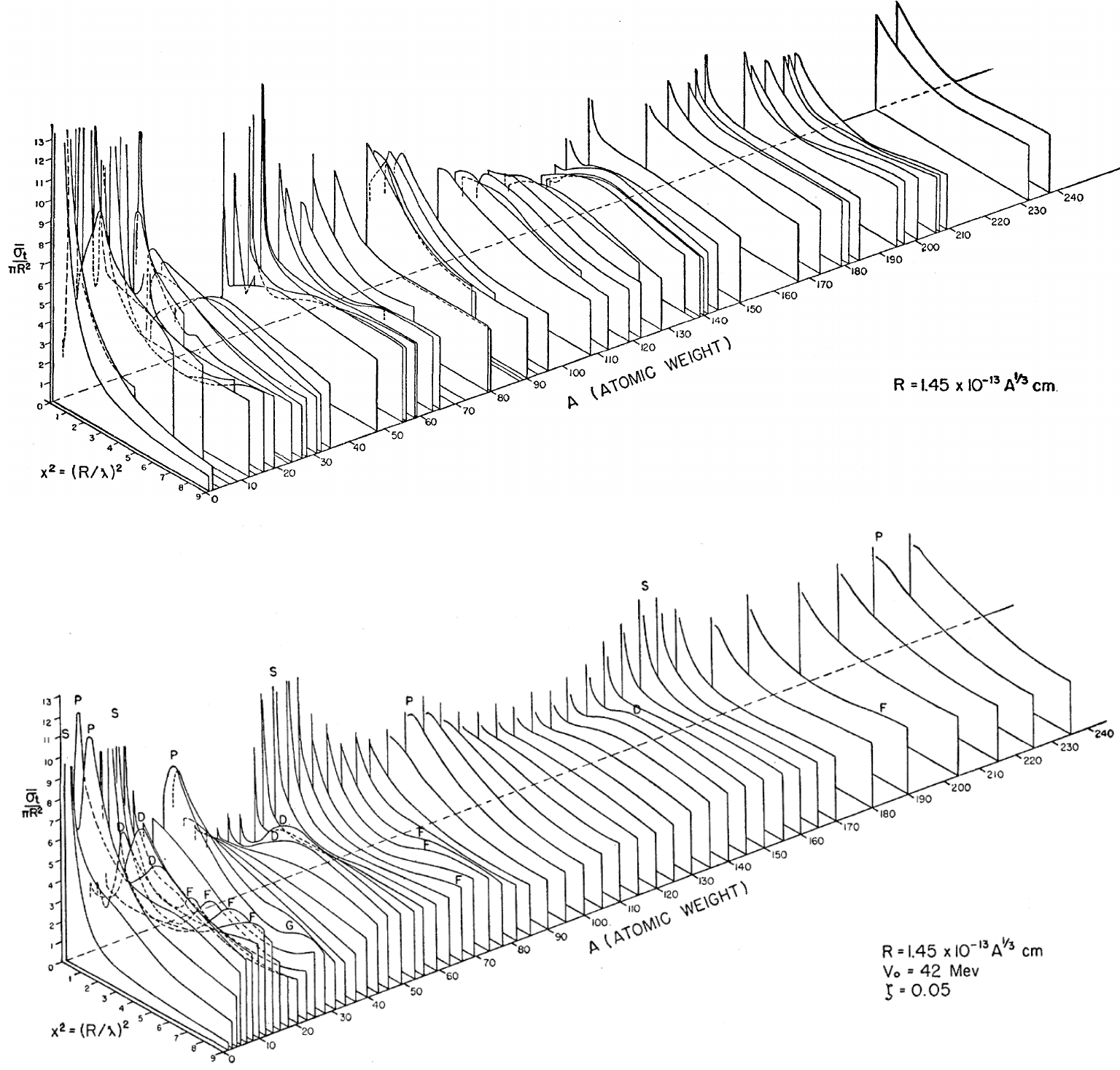}
\caption{Top: experimental neutron total cross section as a function of the mass number $A$ and projectile momentum $p$ expressed in terms of the dimensionless quantity $x^2 = (Rp)^2$. Bottom: same as top, except for the neutron total cross section calculated from the complex single-particle potential $U = -V_0 - i \zeta V_0$ for $r<R$ and $U=0$ otherwise. Reprinted figure with permission from Ref.\ \cite{feshbach54}, copyright (1954) by the
American Physical Society.}
\label{feshfig}
\end{figure}

Original applications of the optical model grew out of the surprising discovery that the total elastic and total reaction cross sections for projectile nucleons on target nuclei are quite smooth as a function of projectile energy and target mass number. This suggests a description in which nucleons scatter from drops of nuclear matter with associated average one-body scattering potential, whose range is roughly proportional to the size of the liquid drop. Since the nucleon can either be scattered elastically or undergo any number of different inelastic reaction processes, the potential should contain both a real and imaginary component \cite{bethe40}, $U(\vec r) = V(\vec r) + i W(\vec r)$, much like the scattering and absorption of light in a dielectric medium is represented by a complex index of refraction. In the center-of-mass frame, the one-body wave equation then takes the form
\begin{equation}
\left ( -\frac{1}{2\mu}\vec \nabla^2 + U(\vec r) - E \right ) \Psi(\vec r) = 0,
\label{schr}
\end{equation}
where $\mu$ is the reduced mass and $U$ is the complex optical potential.
From the time-dependent Sch\"odinger equation and its complex conjugate
\begin{align}
\Psi^*(\vec r, t) \left [ i \frac{\partial \Psi(\vec r, t)}{\partial t} \right ] &= \Psi^*(\vec r, t)\left [ \left ( -\frac{1}{2\mu}\vec \nabla^2 + V(\vec r) +i W(\vec r) \right )\Psi(\vec r,t) \right ] \\
\Psi(\vec r, t) \left [ -i \frac{\partial \Psi^*(\vec r, t)}{\partial t} \right ] &= \Psi(\vec r, t)\left [ \left ( -\frac{1}{2\mu}\vec \nabla^2 + V(\vec r) -i W(\vec r) \right )\Psi^*(\vec r,t) \right ],
\end{align}
one obtains the change in probability density
\begin{equation}
\frac{\partial | \Psi(\vec r, t)|^2}{\partial t} = -\vec \nabla \cdot \vec j(\vec r, t) + 2 W(\vec r)| \Psi(\vec r, t)|^2,
\label{probcurr}
\end{equation}
where the probability current is $\vec j(\vec r,t) = \frac{1}{2i\mu}(\Psi^*(\vec r,t)\nabla \Psi(\vec r,t)-\Psi(\vec r,t)\nabla \Psi^*(\vec r,t))$. One observes that Eq.\ \eqref{probcurr} is the usual continuity equation with the addition of the imaginary part of the optical potential, which for $W(\vec r)<0$ acts as a sink that reduces the elastic scattering flux. In the top panel of Figure \ref{feshfig} is shown the experimental total cross section for neutron projectiles with momentum $p$ (given in terms of the quantity $x = Rp$) on target isotopes with mass number $A$. In the bottom panel of Figure \ref{feshfig} is shown the calculated total cross section for neutron projectiles on target isotopes using a complex one-body potential given by 
\begin{equation}
U = 
\begin{cases}
-V_0 - i \zeta V_0 & r < R \\
0 & r > R.
\end{cases}
\label{sw}
\end{equation}
It is seen that the choices $V_0 = 42$\,MeV, $\zeta = 0.05$, and $R=1.45 \times 10^{-13} A^{1/3}$\,cm produce an overall good reproduction of the experimental data. 

\section{\textit{General form of phenomenological optical potentials}}

The complex square well potential in Eq.\ \eqref{sw} in general gives rise to a much lower reaction cross section compared to experiment \cite{feshbach54,walt55}, since the sharp surface is responsible for greater reflection from the nuclear boundary. Early optical model potentials therefore included a multiplicative Woods-Saxon shape function:
\begin{equation}
U(r) = (V + i W)f(r),
\label{uwws}
\end{equation}
where the parameters $V$ and $W$ are still constant (independent of energy), and $f(r)$ is given by
\begin{equation}
f(r) = \frac{1}{1+e^{(r-R)/a}}
\end{equation}
with $R$ the nuclear radius and $a$ the diffuseness. At low energies ($E \lesssim 50$\,MeV), absorption takes place primarily on the nuclear surface, and therefore it is common to replace the imaginary part of the optical potential in Eq.\ \eqref{uwws} with a surface-peaked form
\begin{equation}
W_D(r) = - 4\, a\, w_D \frac{d}{dr}f(r),
\end{equation}
where the additional factor of $-4a$ is used to enforce the convenient normalization $W_D(R) = w_D$.
The inclusion of real and imaginary one-body spin-orbit interactions
\begin{equation}
U_{so}(r) = (V_{so} + i\, W_{so})\dfrac{1}{r} \dfrac{d}{dr}f(r) \vec l \cdot \vec s,
\end{equation}
where $\vec l$ is the single-particle orbital angular momentum operator and $\vec s$ is the single-particle spin angular momentum operator, allows for a better reproduction of spin observables, such as the vector analyzing power
\begin{equation}
A_y = \frac{1}{P_y}\frac{d\sigma_\uparrow - d\sigma_\downarrow}{d\sigma_\uparrow + d\sigma_\downarrow},
\end{equation}
where $P_y$ is the beam polarization perpendicular to the scattering plane, $d\sigma_\uparrow$ is the differential cross section for spin-up projectiles, and $d\sigma_\downarrow$ is the differential cross section for spin-down projectiles. Finally, for proton scattering one also must include the Coulomb interaction, where the charge distribution of the nucleus is typically assumed to be uniform within the charge radius $R_C$.

Present-day phenomenological optical potentials have the basic structure outlined above, except that the strength functions multiplying the Woods-Saxon distributions can have more complicated energy dependences. For instance, the modern and widely used phenomenological Koning-Delaroche \cite{koning03} global optical potential is written
\begin{eqnarray}
&&\hspace{-.3in} U(r,E) = V_V(r,E) + i W_V(r,E) + i W_D(r,E) \label{phen} \\ \nonumber 
&& + V_{so}(r,E) \vec l \cdot \vec s + i W_{so}(r,E) \vec l \cdot \vec s + V_C(r),
\label{kdpar}
\end{eqnarray}
consisting of energy-dependent real volume, imaginary volume, imaginary surface, real spin-orbit, and imaginary spin-orbit terms as well as the central Coulomb interaction. One notes a few features of the Koning-Delaroche global optical potential to be compared with microscopic optical potentials discussed in subsequent sections. First, the dependence on the projectile energy and on the geometric properties of the target are separated as follows:
\begin{align}
V_V(r,E) &= v_V(E) f(r,R_V,a_V) \\
W_V(r,E) &= w_V(E) f(r,R_V,a_V) \\
W_D(r,E) &= a_D w_D(E) \dfrac{d}{dr}f(r,R_D,a_D) \\
V_{so}(r,E) &= v_{so}(E) \dfrac{1}{r} \dfrac{d}{dr}f(r,R_{so},a_{so}) \\
W_{so}(r,E) &= w_{so}(E) \dfrac{1}{r} \dfrac{d}{dr}f(r,R_{so},a_{so}) \\ 
V_C(r) &= 
\begin{cases} 
      \dfrac{Zze^2}{2R_C}\left ( 3 - \dfrac{r^2}{R_C^2} \right ) & r \leq R_C \\
      \dfrac{Zze^2}{r} & r \geq R_C,
\end{cases}
\end{align}
where $Z$ is the charge of the target and $z$ is the charge of the projectile. The two central volume terms $V_V(r,E)$ and $W_V(r,E)$ are proportional to Woods-Saxon distributions with identical radius and diffuseness parameters. The real and imaginary spin-orbit interactions are proportional to the radial derivative of a single Woods-Saxon distribution whose geometry parameters are allowed to freely vary from those of the central interactions. The imaginary surface contribution $W_D$ is also proportional to the derivative of a Woods-Saxon distribution with independent geometry parameters. The Coulomb interaction is that of a uniformly-charged sphere of fixed radius $R_C$. The energy-dependent multiplicative strength functions for each term have only a few adjustable parameters, and in total the Koning-Delaroche global phenomenological optical potential has 35 parameters.

The first attempts \cite{fernbach49,lelevier52,feshbach54} to model experimental scattering data via complex one-body potentials were purely phenomenological and lacked a clear connection to quantum many-body theory. In the 1950's several different methods were used to derive optical model potentials based on operator projection formalism as well as second quantization techniques borrowed from quantum field theory. These approaches complement calculations of nuclear ground states and excited states using similar methods and have the advantage that they can employ high-precision nuclear forces for a microscopic description of nuclear scattering and reactions.

\section{\textit{Microscopic nuclear forces}}
The description of nucleon-nucleus scattering in terms of the fundamental theory of strong interactions, quantum chromodynamics (QCD), may one day become feasible, however, it almost certainly will not be the most efficient or practical approach for analyzing a wide range of experimental data. Instead, the low-energy realization of QCD, chiral effective field theory \cite{weinberg79} with its explicit nucleon and pion degrees of freedom, is more naturally suited to describe low-energy nuclear phenomena while at the same time incorporating the symmetries and symmetry breaking pattern of QCD. Most modern approaches to constructing nucleon-nucleus optical potentials start from microscopic nuclear forces derived from chiral effective field theory, which includes realistic nuclear microphysics, such as multi-pion exchange processes and three-body forces, that have been shown from ab initio calculations to be crucial for the accurate description of nuclear structure and the saturation phenomenon of nuclear matter. The short-distance details of the nuclear force are not resolved at moderate projectile energies up to $E_{\rm max} \simeq 150-200$\,MeV, and can therefore be encoded in a series of low-energy constants that characterize the strengths of generalized contact interactions. The long-range details of the nuclear force are governed by pion-exchange processes constrained by chiral symmetry. In addition, effective field theory methods allow for the order-by-order expansion of the nuclear force in powers of the low-energy physical scale $q$ divided by the chiral symmetry breaking scale $\Lambda \sim 1$\,GeV. This allows one to quantify the effects of missing physics beyond the truncated order in the effective field theory expansion \cite{wesolowski16,drischler20}. The reader is referred to standard review articles \cite{epelbaum09rmp,machleidt11} for further details.

One of the important goals of modern nuclear reaction theory methods is to elucidate the role of three-body nuclear forces, which in practice can be challenging to implement in microscopic and ab initio frameworks. Whereas two-nucleon forces arise at leading order (LO), $(q/\Lambda)^0$, in the chiral expansion, three-body forces arise at next-to-next-to-leading-order (N2LO), $(q/\Lambda)^3$, in the chiral power counting without explicit $\Delta$-isobar degrees of freedom and at one order lower, $(q/\Lambda)^2$, in the power counting with explicit $\Delta$s. The technically challenging task of including three-body forces in the formalism of scattering theory is often simplified by the use of a normal-ordered Hamiltonian truncated to the two-body level, as will now be demonstrated.

In second quantization, a general three-body force $V_{3N}$ is written
\begin{equation}
V_{3N} = \frac{1}{36} \sum_{123456} \langle 1 2 3 | \bar V_{3N} | 4 5 6 \rangle
\hat a^\dagger_1 \hat a^\dagger_2 \hat a^\dagger_3 \hat a_6 \hat a_5 \hat a_4
\label{v3nsq}
\end{equation}
where $\bar V_{3N}$ is the antisymmetrized three-body force, and $a_i^\dagger$ ($a_i$) is the creation (annihilation) operator for state $| i \rangle$. Normal ordering the three-body
force with respect to, e.g., the ground state of the noninteracting many-body system yields
\begin{eqnarray}
V_{3N} &=&\frac{1}{6} \sum_{ijk}\langle ijk | \bar V_{3N} | ijk \rangle 
+ \frac{1}{2} \sum_{ij1 4} \langle ij1 | \bar V_{3N} | ij4 \rangle :\!\hat a^\dagger_1 \hat a_4\!:  \\ \nonumber
&+& \frac{1}{4}\sum_{i1245}\langle i12 | \bar V_{3N} | i 4 5 \rangle :\!\hat a^\dagger_1 \hat a^\dagger_2 
\hat a_5 \hat a_4\!:  + {1\over 36} \sum_{123456} \langle 1 2 3 | \bar V_{3N} | 4 5 6 \rangle
:\!\hat a^\dagger_1 \hat a^\dagger_2 \hat a^\dagger_3 \hat a_6 \hat a_5 \hat a_4\!:,
\label{nord}
\end{eqnarray}
where the colons around an operator, $:\! \mathcal{\hat O}\!:$, denote normal ordering. The in-medium two-body force 
\begin{equation}
\frac{1}{4}\sum_{i1245}\langle i12 | \bar V_{3N} | i 4 5 \rangle :\!\hat a^\dagger_1 \hat a^\dagger_2 
\hat a_5 \hat a_4\!:
\end{equation}
is effectively obtained by summing the third particle over the filled states in the noninteracting Fermi sea \cite{holt20}:
\begin{equation}
\bar V_{\rm med} = \sum_{s_3 t_3} \int \frac{d^3k_3}{(2\pi)^3} \theta(k_f-k_3) (1-P_{13}-P_{23}) V_{3N},
\label{ddnn}
\end{equation}
where $k_f$ is the Fermi momentum and $P_{ij}$ is the two-body antisymmetrization operator. The operator $(1-P_{12})$ has been absorbed into the definition of the antisymmetrized medium-dependent NN interaction $\bar V_{\rm med}$. In several of the results shown below, chiral effective field theory three-body forces have been implemented using this two-body normal-ordered approximation.

\section{\textit{Feshbach projection formalism}}
\label{ompf}
Feshbach's projection operator method \cite{feshbach58} was one of the first attempts to build a formal connection between the optical model potential for elastic scattering and quantum many-body theory, including the effects of inelastic scattering channels. One starts from the Sch\"odinger equation for the combined projectile + nucleus system:
\begin{equation}
H\Psi = E\Psi,
\end{equation}
where
\begin{equation}
H = H_A(\vec r_1,\dots,\vec r_A) + T_0 + V(\vec r_0,\vec r_1,\dots,\vec r_A)
\end{equation}
with $H_A$ the Hamiltonian for the $A$-particles of the nucleus, $T_0$ the kinetic energy operator for the projectile, and $V$ the potential describing the interaction of the projectile with the $A$ particles of the target nucleus. The exact eigenvalues and eigenstates of the $A$-particle system
\begin{equation}
H_A\Phi_i = \epsilon_i \Phi_i,
\end{equation}
contain in addition to the nuclear ground and excited states also the continuum states in which one or more nucleons are unbound. One then constructs the total wavefunction of the combined system in the distorted wave approximation as
\begin{equation}
\Psi = \sum_{i=0}^{\infty} \phi_i(\vec r_0)\Phi_i(\vec r_1,\dots,\vec r_A).
\label{dwa}
\end{equation}
For simplicity, proper antisymmetrization between the projectile and target is neglected, i.e.,
\begin{equation}
\Psi = {\cal A} \sum_{i=0}^{\infty} \phi_i(\vec r_0)\Phi_i(\vec r_1,\dots,\vec r_A),
\label{dwaa}
\end{equation}
where ${\cal A}$ is the antisymmetrization operator. This may be a reasonable approximation for large projectile energies \cite{takeda55}, where the wavefunction of the incident nucleon overlaps weakly with those of the bound nucleons. Using the orthonormality of the eigenfunctions $\{\Phi_i\}$, one can derive from Eq.\ \eqref{dwa} a set of coupled equations for the amplitudes $\phi_i$:
\begin{equation}
(T_0 + V_{ii} + \epsilon_i - E ) \phi_i = -\sum_{i \ne j} V_{ij} \phi_j,
\label{coupeq}
\end{equation}
where $V_{ij}(\vec r_0) = \langle \Phi_i | V | \Phi_j \rangle$. The goal is to isolate the elastic scattering amplitude $\phi_0$. From the definitions
\begin{align}
\vec U &= \begin{pmatrix}
\phi_1 \\
\phi_2 \\
\vdots
\end{pmatrix},
\end{align}
\begin{equation}
\vec V_0 = (V_{01}, V_{02}, \dots),
\end{equation}
\begin{equation}
\vec H = H_{ij} = T_0\delta_{ij} + V_{ij} + \epsilon_i\delta_{ij} \hspace{.1in} \text{with} \hspace{.1in} i, j > 0 ,
\end{equation}
one can straightforwardly separate Eq.\ \eqref{coupeq} into two channels
\begin{align}
(T_0 + V_{00} - E)\phi_0 &= -\vec V_0 \vec U \label{fp1} \\
(\vec H - E ) \vec U &= -\vec V_0^\dagger \phi_0,
\label{fp2}
\end{align}
where the nucleus ground state energy has been set to $\epsilon_0 = 0$ for convenience.
Formally solving Eq.\ \eqref{fp2} for $\vec U$ and substituting into Eq.\ \eqref{fp1} yields
\begin{equation}
\left ( T_0 + V_{00} + \vec V_0 \frac{1}{E - \vec H + i \eta } \vec V_0^\dagger - E \right ) \phi_0 = 0,
\end{equation}
where the additional $+i\eta$ is included to ensure outgoing waves.
The effective interaction
\begin{equation}
V_{\rm op} = V_{00} + \vec V_0 \frac{1}{E - \vec H + i \eta } \vec V_0^\dagger
\label{vefff}
\end{equation}
can therefore be identified as the optical model potential. Effectively, this amounts to a Schr\"odinger equation for the wavefunction Eq.\ \eqref{dwa} projected onto the ground state
\begin{equation}
P\Psi(\vec r_0, \vec r_1,\dots, \vec r_A) = \phi_0(\vec r_0)\Phi_0(\vec r_1,\dots,\vec r_A),
\end{equation}
where $P$ is the projection operator $P = | \Phi_0 \rangle \langle \Phi_0 |$. Accounting for anti-symmetrization between the projectile and nucleons in the target nucleus leads to a modified projection operator equation:
\begin{equation}
P\Psi(\vec r_0, \vec r_1,\dots, \vec r_A) = {\cal A} \phi_0(\vec r_0)\Phi_0(\vec r_1,\dots,\vec r_A)
\end{equation}
with a more complicated associated projection operator. For additional details the reader is referred to Ref.\ \cite{feshbach62}.


\section{\textit{Optical potential at high energies: multiple scattering theory}}
\label{omphe}

At sufficiently high energies, the scattering of a projectile nucleon on a target nucleus can be approximated by the superposition of scattering waves generated by each of the constituent nucleons in the target acting independently. In the lowest-order approximation, each of these scatterings is therefore treated as a two-body collision, and more complicated many-body effects such as Pauli blocking and three-body forces can be neglected. The total transition amplitude is therefore written in terms of the individual transition amplitudes for free-space nucleon-nucleon scattering, where one also needs to specify the momentum distribution of nucleons in the nucleus. The projectile wave incident on each of the constituent nucleons is then a combination of a plane wave and all the waves generated from each of the nucleons in the nucleus. One anticipates a modified Schr\"odinger equation of the form
\begin{equation}
\left ( H_0 - E + \int \rho(\vec r)T(\vec r)d\vec r \right ) \langle \psi \rangle = 0,
\label{ms1}
\end{equation}
where $\rho(\vec r)$ is the density of scatters and $T(\vec r)$ is the $T$-matrix for two-body scattering of the incident projectile and scattering nucleon located at position $\vec r$, and $\langle \psi \rangle$ is the average total wave.

The goal will be to derive an equation for the optical model potential in terms of two-body scattering amplitudes. As a necessary condition to apply multiple scattering theory, the projectile wavelength $\lambda \sim 1/k$ should be small compared to the interparticle separation distance $r_0$ between nucleons in the nucleus:
\begin{equation}
k \gg \frac{1}{r_0}.
\label{mstc}
\end{equation}
In this case, the wave propagating from one scatterer to a nearby scatter is just the asymptotic form, namely an outgoing wave multiplied by the scattering amplitude. Eq.\ \eqref{mstc} requires that the incident nucleon have sufficiently large energy, and conservatively one may identify the applicability region to projectile energies $E \gtrsim 70$\,MeV (additional details and analyses using chiral effective field theory interactions can be found in Refs.\ \cite{burrows19,burrows20}).

One begins from the potential energy between projectile nucleon at $\vec r_0$ and the $A$ target nucleons in the nucleus:
\begin{equation}
V = \sum_{i=1}^A v(\vec r_0, \vec r_i).
\label{vsum}
\end{equation}
From this one defines the transition matrix $T$ by
\begin{equation}
T = V + V\frac{1}{E-H_0+i\eta}T.
\end{equation}
Formally this equation is similar to that of the free-space $T$-matrix but includes bound and continuum states of the $A$-particle nucleus:
\begin{equation}
\langle n^\prime \vec k^\prime | T | \vec k n \rangle = 
\langle n^\prime \vec k^\prime | V | \vec k n \rangle + 
\sum_m \int \frac{\langle n^\prime \vec k^\prime | V | \vec p m \rangle \langle m \vec p | T | \vec k n \rangle}{E - E_m(\vec p) + i \eta} d\vec p,
\label{kmt1}
\end{equation}
where the intermediate-state energy is given as the sum of the nucleus energy $\epsilon_m$ and the kinetic energy $T_0(\vec p)$ of the projectile. The evaluation of Eq.\ \eqref{kmt1} can be simplified by noting that for antisymmetrized states of the nucleus, the $A$ independent terms in Eq.\ \eqref{vsum} are identical:
\begin{equation}
T = Av \left( 1 + \frac{1}{E-H_0+i\eta} T \right ).
\label{T1}
\end{equation}
Note that antisymmetrization between the projectile and target nucleus has been neglected.

From Eq.\ \eqref{T1} one sees that the first-order approximation is just the sum of the $A$ scattering amplitudes in the {\it Born approximation}. However, it is desired to develop an expansion formalism in which the leading term is the sum of the $A$ {\it full scattering amplitudes}. For this one introduces a new scattering operator
\begin{equation}
\tau = v \left ( 1 + \frac{1}{ E - H_0 + i \eta} \tau \right )
\label{tau1}
\end{equation}
and seek to replace $v$ in Eq.\ \eqref{T1} with $\tau$, which accounts for multiple scattering. This can be achieved by solving Eq.\ \eqref{tau1} for $v$ and substituting it into Eq.\ \eqref{T1}:
\begin{align}
v &= \left ( 1 + \tau \frac{1}{E - H_0 + i\eta} \right )^{-1}\tau \\
T &= \left ( 1 - ( A - 1 ) \tau \frac{1}{E - H_0 + i \eta} \right )^{-1} A \tau \\
T &= A\tau + (A-1)\tau\frac{1}{E - H_0 + i \eta}T.
\end{align}
Defining $T^\prime = (A-1) T / A$, one obtains
\begin{equation}
T^\prime = (A-1)\tau \left ( 1 + \frac{1}{E - H_0 + i \eta} T^\prime \right ),
\end{equation}
where the term $V^{(e)} \equiv (A-1)\tau$ is identified as the effective potential. In fact, from this modified scattering matrix equation and effective potential, one can rederive the optical potential using the Feshbach projection formalism of the previous section simply replacing $V \longrightarrow V^{(e)}$. One arrives at the result
\begin{equation}
V_{\rm op}^{(e)} = V_{00}^{(e)} + \vec V_0^{(e)} \frac{1}{E - \vec H^{(e)} + i \eta } \vec V_0^{(e)\dagger}.
\end{equation}

In practice one applies two further approximations. First, at high energies it may be reasonable to neglect the effect from excited nuclear states on the elastic scattering cross section:
\begin{equation}
V_{\rm op}^{(e)} \simeq V_{00}^{(e)} = (A-1)\langle \phi_0 | \tau | \phi_0 \rangle.
\end{equation}
In other words, only multiple scattering effects are taken into account. Second, the many-body scattering matrix $\tau$ is replaced by the free-space NN scattering matrix $T_{NN}$ (referred to as the impulse approximation):
\begin{equation}
V_{\rm op}^{(e)} \simeq (A-1)\langle \phi_0 | T_{NN} | \phi_0 \rangle,
\end{equation}
which is nonlocal and depends on the energy associated to the $NN$ $T$-matrix. Thus, one may write the optical potential in the form
\begin{equation}
U_i(\vec q, \vec q^\prime) = \frac{A-1}{A} \sum_{sj}\! \int d \vec p\, \langle \vec q^\prime, \vec p -\!\frac{\vec q^\prime -\vec q}{2}, s, j | T_{ij}(\epsilon) | \vec q, \vec p +\! \frac{\vec q^\prime -\vec q}{2}, s, j \rangle \rho_j^s(\vec q, \vec q^\prime; \vec p),
\label{kmt1}
\end{equation}
where $i$ ($j$) labels the isospin projection of the projectile (struck nucleon in the target), $\rho_j^s(\vec q, \vec q^\prime; \vec p)$ is the one-body density matrix of the target associated with particles having isospin projection $j$ and spin projection $s$.
One may invoke several approximations \cite{picklesimer84} to rewrite Eq.\ \eqref{kmt1} in the form of the $NN$ $T$-matrix multiplied by the nuclear density, e.g., 
\begin{equation}
U_i(\vec q, \vec q^\prime; \epsilon) \sim \frac{A-1}{A} \sum_j T_{ij}(\vec q, \vec q^\prime;\epsilon) \rho(| \vec q^\prime - \vec q|).
\end{equation}
More recently, there have been efforts \cite{burrows19,burrows20} to improve upon several of these approximations, including the use of ab initio one-body densities as well as explicit account of the spin of the struck nucleon in the target.

\begin{figure}[t]
\begin{center}
\includegraphics[width=10cm]{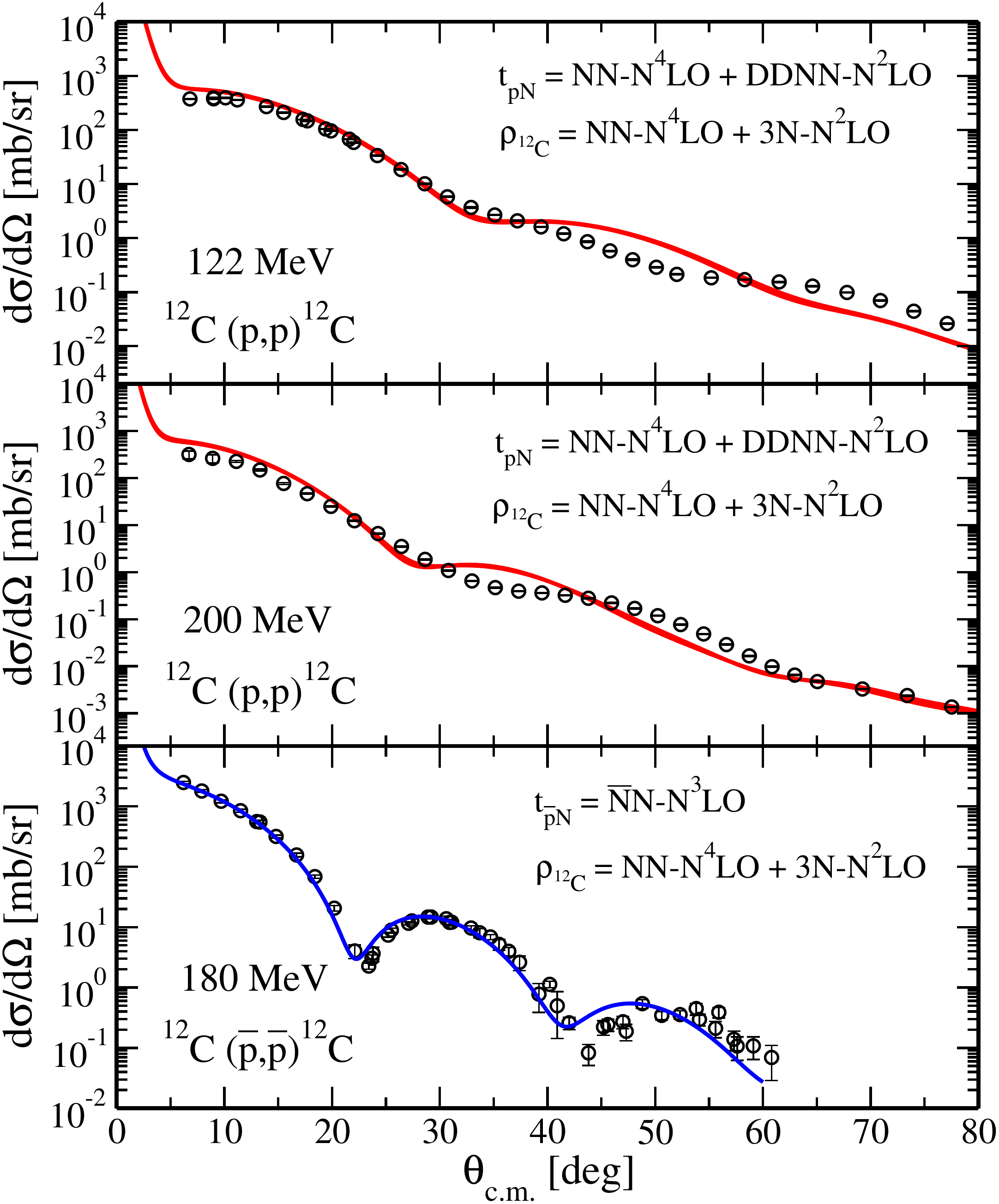}
\caption{ (Upper two panels) Differential cross sections as a function of the center-of-mass scattering angle for elastic proton scattering off $^{12}$C at 122 MeV and 200 MeV. The band shows the result obtained with the $t_{pN}$ matrix computed with the $pN$ chiral interaction of Ref.~\cite{Entem:2017gor} supplemented by a density-dependent $NN$ interaction~\cite{holt2010} (with $ 0.08$ fm$^{-3} \le \rho \le 0.13$ fm$^{-3}$) and the one-body nonlocal density matrix computed with the NCSM method~\cite{barrett2013} using $NN$~\cite{Entem:2017gor} and $3N$~\cite{Navratil2007,Gysbers2019} chiral interactions. (Lower panel) Differential cross section as a function of the center-of-mass scattering angle for elastic antiproton scattering off $^{12}$C at 180 MeV. The result is obtained with the $t_{\bar{p}N}$ matrix computed with the $\bar{p}N$ chiral interaction of Ref.~\cite{dai2017} and the one-body nonlocal density matrix computed with the NCSM method using $NN$~\cite{Entem:2017gor} and $3N$~\cite{Navratil2007,Gysbers2019} chiral interactions.
\label{fig_12C_sigma}}
\end{center}
\end{figure}

In the top two panels of Figure \ref{ms1} is shown the differential elastic scattering cross sections for proton projectiles on a $^{12}$C target at energies $E=122$\,MeV and $200$\,MeV computed within a multiple-scattering formalism \cite{vorabbi20} compared to experimental data. The proton-nucleon $T$-matrix, $t_{pN}$, is derived from nucleon-nucleon interactions at N4LO in the chiral power counting together with a density-dependent two-body force \cite{holt2010} derived from the chiral effective field theory three-body force at N2LO. The target one-body nonlocal density matrix entering into the derivation of the optical potential is obtained from the same nucleon-nucleon interaction and the full N2LO chiral three-body force \cite{Navratil2007,Gysbers2019}. One finds overall quite good agreement with the measured data across most scattering angles and for the two energies considered (see also Refs.\ \cite{burrows19,burrows20} for related work). In the lower panel of Fig.\ \ref{ms1} is shown the elastic scattering cross section for antiprotons on a $^{12}$C target at energy $E=180$\,MeV. The theory results are obtained using the antiproton-nucleon $T$-matrix computed from the $\bar pN$ chiral interaction at N3LO \cite{dai2017} together with the same $^{12}$C one-body nonlocal density matrix used in computing the proton scattering cross sections above. Again, the multiple scattering theory is able to predict well the measured cross sections.

\section{\textit{Optical potential from Green's function theory}}

While formally satisfying, the Feshbach construction of the effective Hamiltonian operator that projects onto the elastic scattering channel (and projects out the inelastic channels) can be difficult to implement in practice. An alternative approach begins from a description of single-particle motion based on the one-body Green's function \cite{bell59,escher02,dickhoff18}:
\begin{equation}
G(\vec r, t; \vec r^\prime, t^\prime) = -i \langle \Psi^A_0 | \hat T [ \hat a(\vec r,t)_H \hat a^\dagger(\vec r^\prime,t^\prime)_H ] | \Psi^A_0 \rangle,
\end{equation}
which is the ground-state expectation value of the time-ordered product of nucleon creation/annihilation operators in the Heisenberg representation (denoted by the subscript $H$). The Green's function characterizes the propagation of particles or holes added to a many-body ground state. For instance, in momentum space the particle part of the Green's function $G_p(\vec k;t)$ describes the probability amplitude of finding a particle at time $t>0$ with momentum $\vec k$ if it was added to the ground state of the many-body system at time $t=0$ with the same momentum $\vec k$. For a Hamiltonian invariant under time translation, one has the Fourier transformed Green's function
\begin{align}
G(\vec r, \vec r^\prime;E) &= \langle \Psi^A_0 | a(\vec r) \frac{1}{E-(H-E_0^A)+ i\eta} a^\dagger(\vec r^\prime) | 
\Psi^A_0 \rangle \\
& + \langle \Psi^A_0 | a^\dagger(\vec r^\prime) \frac{1}{E+(H-E_0^A)- i\eta} a(\vec r) | 
\Psi^A_0 \rangle,
\end{align}
where $H |\Psi^A_0 \rangle = E_0^A |\Psi^A_0 \rangle$. Inserting a complete set of eigenstates for the $A+1$ and $A-1$ many-body systems, one obtains the so-called Lehmann representation
\begin{equation}
G(\vec r, \vec r^\prime;E) = \sum_n \frac{f_n(\vec r)f^*_n(\vec r^\prime)}{E-(E_n^{A+1}-E_0^A)+ i \eta} + \sum_n \frac{g_n(\vec r)g^*_n(\vec r^\prime)}{E+(E_n^{A-1}-E_0^A)- i \eta},
\end{equation}
where $f_n(\vec r) = \langle \Psi^A_0 | a(\vec r) | \Psi^{A+1}_n \rangle$ and $g_n(\vec r) = \langle \Psi^{A-1}_n | a(\vec r) | \Psi^A_0 \rangle$ are overlap functions related to the probability for transition to bound or scattering states of the $A-1$ or $A+1$ systems. One can show that the particle-hole Green's function satisfies the equation motion
\begin{equation}
\left ( E + \frac{\nabla_r^2}{2M} \right ) G(\vec r, \vec r^\prime; E) - \int \Sigma(\vec r, \vec r^{\prime\prime}; E) G(\vec r^{\prime\prime}, \vec r^\prime; E) d \vec r^{\prime\prime} = \delta(\vec r - \vec r^\prime),
\label{eom}
\end{equation}
where the nucleon self energy $\Sigma$ has been introduced and is defined by
\begin{equation}
G(\vec r, \vec r^\prime; E) = G_0(\vec r, \vec r^\prime; E) + \int d\vec y \int d\vec y^\prime G_0(\vec r, \vec y; E) \Sigma(\vec y, \vec y^\prime; E) G(\vec y^\prime, \vec r^\prime; E).
\end{equation}
Substituting the Lehmann representation of the Green's function into Eq.\ \eqref{eom} yields for the single-particle amplitudes
\begin{equation}
\left ( (E_n^{A+1} - E_0^A) + \frac{\nabla_r^2}{2M} \right ) f_n(\vec r) - \int \Sigma(\vec r, \vec r^{\prime}; E_n) f_n(\vec r^\prime) d \vec r^{\prime} = 0.
\label{wave}
\end{equation}
Therefore, one sees that the self energy $\Sigma(\vec r, \vec r^{\prime}; E_n)$ acts as an energy-dependent and non-local single-particle potential in the wave equation Eq.\ \eqref{wave} for the amplitude $f_n$. Bell and Squires showed \cite{bell59} that the spectrum of scattering states for the Hamiltonian
\begin{equation}
H(\vec r, \vec r^\prime) = -\frac{\nabla_r^2}{2M} \delta(\vec r - \vec r^\prime) + \Sigma(\vec r, \vec r^{\prime}; E)
\end{equation}
is exactly equal to the overlap functions associated with the elastic scattering channel, and hence $\Sigma(\vec r, \vec r^{\prime}; E)$ is identified as the optical model potential. It is worth highlighting that the above equations for the Green's function and self energy are expressed in terms of single-particle coordinates $\vec r$ and $\vec r^\prime$, rather than relative coordinates with respect to the target's center of mass. The application of the Green's function formalism in ab initio nuclear reaction theory therefore requires additional approximations since in general the center-of-mass motion is not completely suppressed \cite{idini19}.

\begin{figure}[t]
\begin{center}
\raisebox{0.1cm}{\includegraphics[width=5.7cm]{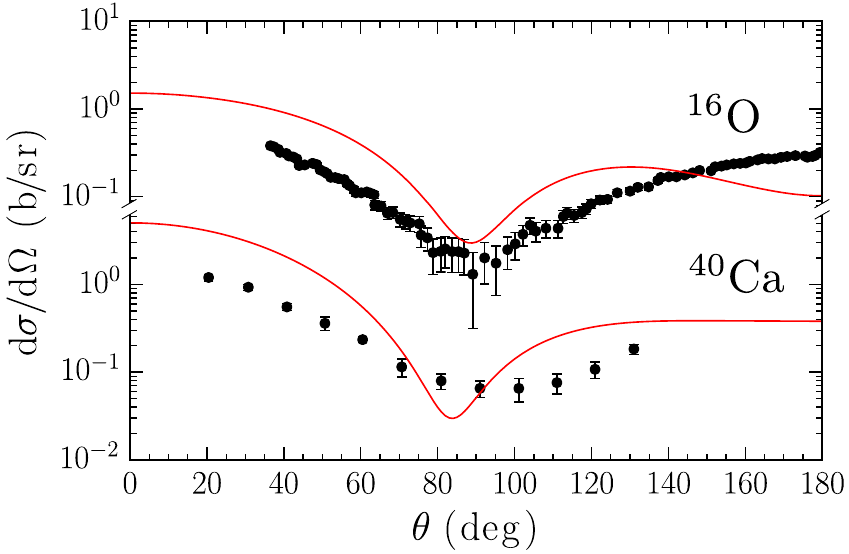}}
\includegraphics[width=5.7cm]{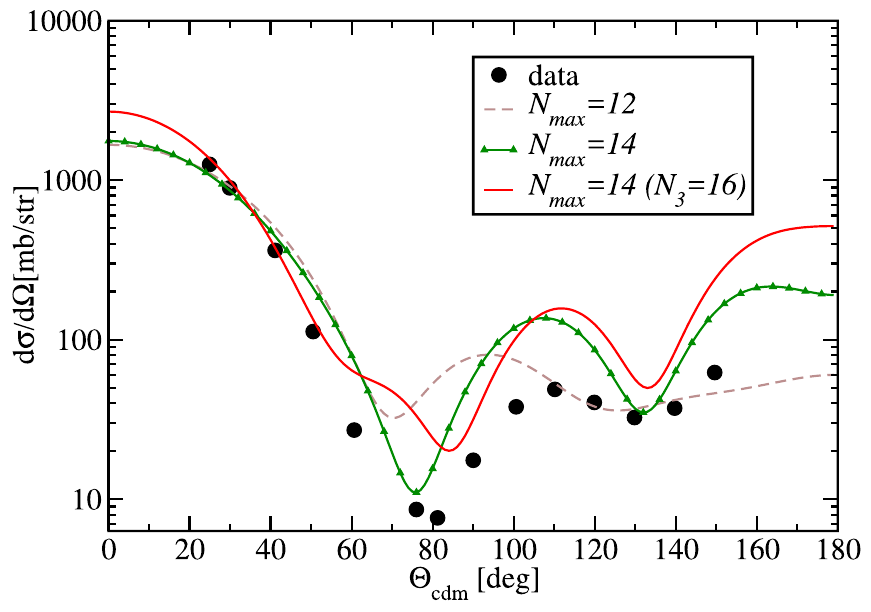}
\caption{ (Left panel) Differential elastic scattering cross section as a function of the center-of-mass scattering angle for neutron scattering off $^{16}$O ($^{40}$Ca) at 3.286 (3.2) MeV from $NNLO_{\rm sat}$ (red curves) compared to data. Reprinted figure with permission from Ref.\ \cite{idini19}, copyright (2019) by the American Physical Society. (Right panel) Differential elastic scattering cross section as a function of scattering angle for a neutron projectile at 5.17 MeV on a $^{40}$Ca target. Coupled-cluster calculations using the $NNLO_{\rm sat}$ chiral potential are shown for different model spaces of the 2N force ($N_{max}$) and 3N force ($N_3$) and compared to data. Reprinted figure with permission from Ref.\ \cite{rotureau18}, copyright (2018) by the American Physical Society.
\label{ai} }
\end{center}
\end{figure}

The Green's function approach has the advantage that a perturbation expansion for the irreducible self energy $\Sigma(\vec r, \vec r^{\prime}; E)$ is straightforward to develop. Additionally, antisymmetrization between the nucleon projectile and target nucleus is naturally accounted for in calculations of the self energy using the standard techniques of second quantization. Note, however, that translational invariance may not be automatically satisfied, and recent work \cite{johnson17,johnson19} has sought to develop a modified translationally-invariant second-quantized theory of optical model potentials. Within the Green's function formalism, the self energy for states below the Fermi surface can also be linked to the shell model potential, however, to include consideration of the imaginary part it is more convenient to define the so-called ``mass operator'' ${\cal M}$ as
\begin{equation}
{\cal M}(k;E) =
\begin{cases}
\Sigma(k;E) & E > E_F \\
\Sigma(k;E)^* & E < E_F.
\end{cases}
\end{equation}
The mass operator has the property that it is an analytical function in the upper-half plane.

Green's function theory has been used in ab initio many-body calculations to construct optical potentials and corresponding differential elastic scattering cross sections for low-energy neutrons on selected doubly-magic nuclei. In the left panel of Fig.\ \ref{ai} is shown as the red curves the elastic scattering cross sections of 3.286 MeV neutrons on $^{16}$O and 3.2 MeV neutrons on $^{40}$Ca computed from optical model potentials generated within self-consistent Green's function theory using the high-precision chiral 2N + 3N interaction $NNLO_{sat}$ \cite{ekstrom15}. One sees that in general the local minima are reproduced fairly well, but overall the elastic scattering results are overestimated compared to data (black dots), indicating that the absorptive strength of the ab initio optical potential is too small. This is likely the result of missing configurations, such as 3-particle--2-hole states, that at present cannot be computed within the self-consistent Green's function formalism and that would open additional inelastic scattering channels. In the right panel of Fig.\ \ref{ai} is shown the elastic scattering cross section of 5.17\,MeV neutron projectiles on $^{40}$Ca as a function of scattering angle computed within a coupled cluster Green's function formalism \cite{rotureau18} (colored curves) and compared to experimental data (black dots). Results are shown for several different values of the model-space parameter $N_{max}$ for which all shells with principal quantum number $n$ and orbital angular momentum quantum number $l$ satisfying $2n+l \leq N_{max}$ are included. In general, the parameter $N_3$ characterizing the three-nucleon force model-space cutoff is taken the same ($N_3 = N_{max}$), except for the results shown by the red line for which $N_{max}=14$ and $N_3=16$, highlighting the important role of three-body forces. One observes that achieving converged results is challenging within the coupled cluster Green's function approach, though in general the theoretical cross section predicts correctly the angles of the scattering local minima as well as the scattering cross section at forward angles.

\section{\textit{Green's function theory for homogeneous nuclear matter}}

An alternative approach \cite{jeukenne76} to the Green's function formalism outlined in the previous section starts by calculating the nucleon self energy in homogeneous infinite nuclear matter at varying isoscalar density $n = n_n + n_p$ and isospin asymmetry $\delta_{np} = (n_n - n_p ) / (n_n + n_p )$. In terms of the proton and neutron Fermi momenta $k_f^p = (3 \pi^2 (1 - \delta_{np}) n / 2)^{1/3}$ and $k_f^n = (3 \pi^2 (1 + \delta_{np}) n / 2)^{1/3}$, the nucleon optical potential in infinite homogeneous matter is then written:
\begin{eqnarray}
U_p(E;k_f^p,k_f^n) &=& V_p(E;k_f^p,k_f^n) + i W_p(E;k_f^p,k_f^n), \nonumber \\ 
U_n(E;k_f^p,k_f^n) &=& V_n(E;k_f^p,k_f^n) + i W_n(E;k_f^p,k_f^n).
\label{omp}
\end{eqnarray}
To obtain a nucleon-nucleus optical potential, the nucleon self energy in homogeneous matter is folded with the target nucleus density distribution in the local density approximation:
\begin{equation}
V(E;r) + i W(E;r) = V(E;k_f^p(r),k_f^n(r)) + i W(E;k_f^p(r),k_f^n(r)).
\end{equation}
The nucleon self-energy in an infinite system needs to be computed only once and then can be applied to any finite system for which the density-distribution is known. The nuclear matter approach is therefore ideal for constructing microscopic {\it global optical potentials} capable of accommodating a wide range of target nuclei and projectile energies.

\begin{figure}[t]
\centering
\includegraphics[scale=0.5]{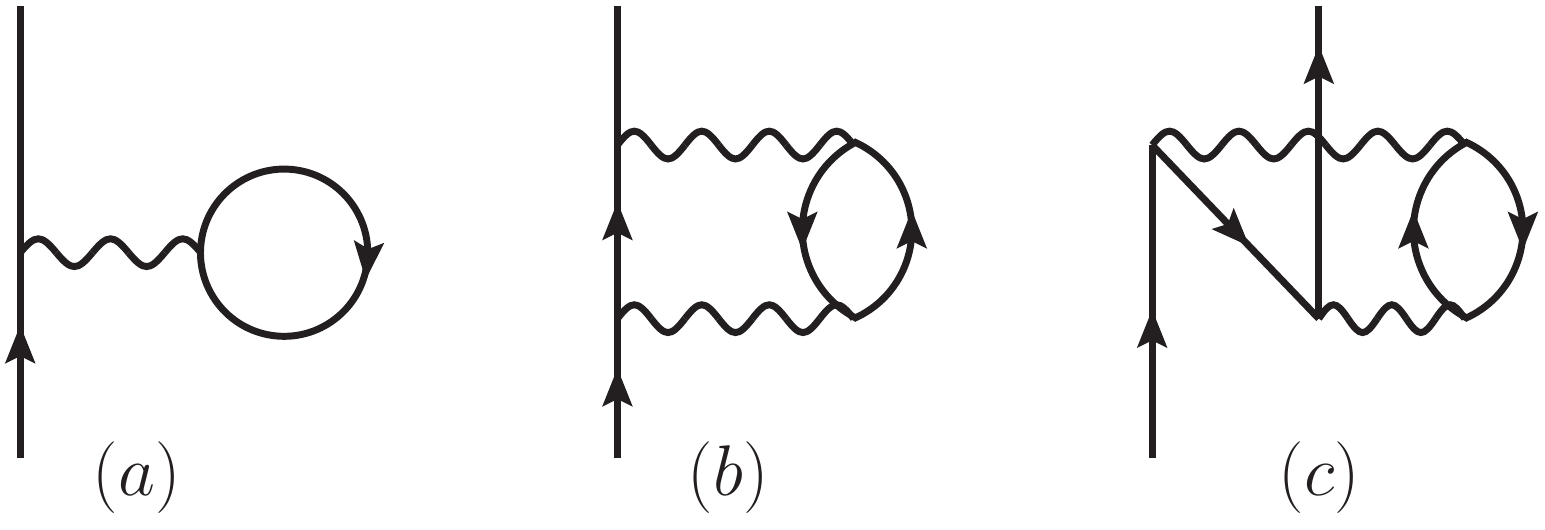}
\caption{First- and second-order perturbation theory diagrams contributing to the nucleon self energy. The solid lines with up (down) arrows represent particle (hole) states, and the wavy line represents the nucleon-nucleon interaction.}
\label{se}
\end{figure}

In Figure \ref{se} is shown the first- and second-order perturbative contributions to the nucleon self energy, considering only the presence of two-body forces. Upward (downward) going solid lines represent nucleon states with momenta above (below) the Fermi surface, and the wavy line represents the antisymmetrized nuclear two-body interaction $\bar V = (1-P_{12}) V$ with $P_{12}$ the permutation operator
\begin{equation}
P_{12} = \left ( \frac{1+ \vec \sigma_1 \cdot \vec \sigma_2}{2} \right ) \left ( \frac{1+ \vec \tau_1 \cdot \vec \tau_2}{2} \right ),\,\, \vec k_1 \longleftrightarrow \vec k_2.
\end{equation}
Diagram (a) is the first-order (or Hartree-Fock) contribution, which in isospin-symmetric nuclear matter is given by
\begin{equation}
\Sigma^{(1)}_{2N}(k)=\sum_{1} \langle \vec{k} \: \vec{h}_1 s s_1 t t_1 | \bar{V}_{2N} | \vec{k} \: \vec{h}_1 s s_1 t t_1 \rangle n_1 ,
\label{eq:2}
\end{equation}
where $\vec k, s, t$ are the momentum, spin, and isospin of the projectile, $n_1$ is the occupation probability $\theta(k_f - h_1)$ for a filled state with momentum $\vec h_1$ below the Fermi surface and the summation is over intermediate-state momenta $\vec h_1$, spins $s_1$, and isospins $t_1$. For example, if one considers a nucleon propagating in symmetric nuclear matter at density $n=2k_f^3/3\pi^2$ and interacting through a scalar-isoscalar boson exchange interaction
\begin{equation}
V = \frac{g^2}{m^2+q^2},
\end{equation}
with $\vec q$ the momentum transfer, then the corresponding direct (D) and exchange (E) terms are written \cite{holt13}
\begin{align}
\Sigma^{(1)}_D &= 4 \int \frac{d^3h_1}{(2\pi)^3} \frac{g^2}{m^2}\theta(k_f - h_1) = \frac{2g^2}{3\pi^2}\frac{k_f^3}{m^2}, \\
\Sigma^{(1)}_E &= \int \frac{d^3h_1}{(2\pi)^3} \frac{g^2}{m^2+(\vec k - \vec h_1)^2}\theta(k_f - h_1) = \frac{g^2m}{4\pi^2} \left (\frac{k_f}{m} - \arctan{ \left (\frac{k_f}{m} + \frac{k}{m} \right ) }  \right . \\
&- \left . \arctan{ \left (\frac{k_f}{m} - \frac{k}{m} \right ) } + \frac{m^2 + k_f^2 - k^2}{4mk} \ln{\frac{m^2 + (k_f + k)^2}{m^2 + (k_f - k)^2} } \right ).
\end{align}
One sees that the direct term is just a constant and hence local in coordinate space, whereas the exchange term depends on the magnitude of the projectile momentum $k$. Moreover, both contributions have no imaginary part and do not explicitly depend on the energy of the projectile. However, employing the relation
\begin{equation}
\label{eq:1}
E(k)=\frac{k^2}{2M} + \Sigma^{(1)}(k),
\end{equation}
one can interchange the momentum-dependent (i.e., spatially nonlocal) mean field with an energy-dependent local mean field.

The second-order perturbative contributions to the nucleon self energy in symmetric nuclear matter read
\begin{equation}
\label{sig2a}
\Sigma^{(2a)}_{2N}(k;E) = \frac{1}{2} \sum_{123} \frac{|\langle \vec{p}_1 \vec{p}_3 s_1 s_3 t_1 t_3 | \bar{V}_{2N} | \vec{k} \vec{h}_2 s s_2 t t_2 \rangle|^2}{E+\epsilon_2-\epsilon_1-\epsilon_3+i\eta} \bar{n}_1 n_2 \bar{n}_3,
\end{equation}
\begin{equation}
\label{sig2b}
\Sigma^{(2b)}_{2N}(k;E) = \frac{1}{2} \sum_{123} \frac{|\langle \vec{h}_1 \vec{h}_3 s_1 s_3 t_1 t_3 | \bar{V}_{2N} | \vec{k} \vec{p}_2 s s_2 t t_2 \rangle|^2}{E+\epsilon_2-\epsilon_1-\epsilon_3-i\eta}   n_1 \bar{n}_2 n_3,
\end{equation}
where the occupation probability for particle states above the Fermi momentum is $\bar n_i = \theta(k_i-k_f)$. Hence, one sees that the effective optical potential for elastic scattering is influenced by the excited states of the combined system, which serve as possible virtual excitation modes contributing to the scattering amplitude. In contrast to the Hartree-Fock contribution, the second-order contributions $\Sigma^{(2a)}$ and $\Sigma^{(2b)}$ are in general energy-dependent and complex. Contribution $\Sigma^{(2a)}$ is complex only for states above the Fermi surface, including positive-energy scattering states, while contribution $\Sigma^{(2b)}$ is complex only for states below the Fermi surface. In the latter case, the imaginary part describes the width of the state produced in a knockout or transfer reaction that leaves behind a hole vacancy in the target nucleus.

In Eqs.\ \eqref{sig2a} and \eqref{sig2b}, the single-particle energies $E$ and $\epsilon$ should be computed self-consistently according to 
\begin{equation}
E(k) = \frac{k^2}{2M} + \Sigma(k;E(k)).
\label{dyson1}
\end{equation}
In the context of elastic scattering, the energy $E$ in Eq.\ \eqref{dyson1} is real and therefore the momentum $k=k_r+ik_i$ of the propagating nucleon is in general complex, since $\Sigma(k;E(q)) = V(k;E) + i W(k;E)$ is also complex. Since the imaginary part is usually several times smaller than the real part, in practice it is convenient to make the approximation:
\begin{equation}
E = \frac{k_0^2}{2M} + V(k_0;E),
\label{dyson2}
\end{equation}
which defines $k_0$.  Considering both $k_i/k_r$ and $W(k;E)/V(k;E)$ as small parameters and expanding Eq.\ \eqref{dyson1} to first order in both quantities, one obtains
\begin{equation}
E = \frac{k_0^2}{2M}+V(k_0;E) + \left ( \frac{k_0}{M} + \left . \frac{\partial V}{\partial k} \right |_{k_0} \right ) (k_r + ik_i - k_0) + iW(k_0;E)
+ \cdots.
\end{equation}
It follows that
\begin{align}
k_0 = k_r &= \sqrt{2M(E-V(k_r;E))} \\
k_i &= -W(k_r;E) \left ( \frac{k_r}{M} + \left . \frac{\partial V}{\partial k} \right |_{k_r} \right )^{-1} = -\frac{M}{k_r}W
\left ( 1+\frac{M}{k_r} \left . \frac{\partial V}{\partial k} \right |_{k_r} \right )^{-1}.
\label{kri1}
\end{align}
On the other hand, from the Schr\"odinger equation \eqref{schr} for a complex potential $U(r) = V(r) + i W(r)$, one obtains the wavevector solution
\begin{equation}
k = k_r + i k_i = \sqrt{2M(E-V-iW)}
\end{equation}
with real and imaginary components
\begin{align}
k_r &= \sqrt{M(E-V)+M\sqrt{(E-V)^2+W^2}} \simeq \sqrt{2M(E-V)} \\
k_i &= -\frac{M}{k_r}W.
\label{kri2}
\end{align}
Comparing Eqs.\ \eqref{kri1} and \eqref{kri2}, one finds that in order to compare a microscopic optical potential derived under the assumption Eq.\ \eqref{dyson2}, one must include an additional factor $\left ( 1+ \frac{M}{k_r} \left . \frac{\partial V}{\partial k} \right |_{k_r} \right )^{-1}$ for the imaginary part that arises from the inherent spatial nonlocality (momentum dependence) of the optical potential \cite{negele81,fantoni81}.

The nonlocality of the optical potential mean field is directly related to the nucleon effective mass \cite{Li18} in a nuclear medium:
\begin{equation}
\frac{dE}{dk} = v = \frac{k}{M^*} = \frac{k}{M} + \frac{\partial V}{\partial k} + \frac{\partial V}{\partial E}\frac{dE}{dk},
\end{equation}
where
\begin{equation}
\frac{M^*}{M} = \left ( 1 + \frac{M}{k}\frac{\partial V}{\partial k} \right )^{-1} \left ( 1 - \frac{\partial V}{\partial E} \right ) \equiv \frac{M_k^*}{M} \frac{M_E^*}{M}
\end{equation}
depends on both the spatial nonlocality and temporal nonlocality (energy dependence) of the optical potential. The $k$-dependent effective mass is known to be roughly $M_k^*/M \simeq 0.7$ \cite{friedman81}, and therefore the imaginary optical potential computed from microscopic many-body theory under the assumption Eq.\ \eqref{dyson2} should be reduced when accounting for nonlocality effects. In addition, the nuclear mean free path $\lambda$ characterizes the distance over which the flux of nucleons decreases by a factor $e$ and is related to the imaginary wavevector $k_i$ via 
\begin{equation}
\lambda = \frac{1}{2k_i} = -\frac{k_r}{2M_k^*W(k_r;E)},
\end{equation}
and therefore nonlocality effects increase the nucleon mean free path.

The nuclear matter approach as defined above has several obvious shortcomings. First, the spin-orbit force is proportional to the gradient of the potential and therefore vanishes in a homogeneous medium. In the past, it was common to supplement the microscopic central components of the optical potential with a phenomenological spin-orbit interaction. More recently, in Ref.\ \cite{whitehead19} it was suggested to use the density matrix expansion \cite{negele72,gebremariam10} to obtain the leading-order Hartree-Fock approximation to the spin-orbit interaction in nuclei starting from the same nuclear two-body and three-body forces used to compute the nucleon self energy. Such a spin-orbit interaction is constructed at the Fermi surface (rather than at positive scattering energies) and has no energy dependence, but nevertheless has a magnitude similar to that from phenomenology \cite{whitehead19}.

\begin{figure}[t]
\centering
\includegraphics[scale=0.83]{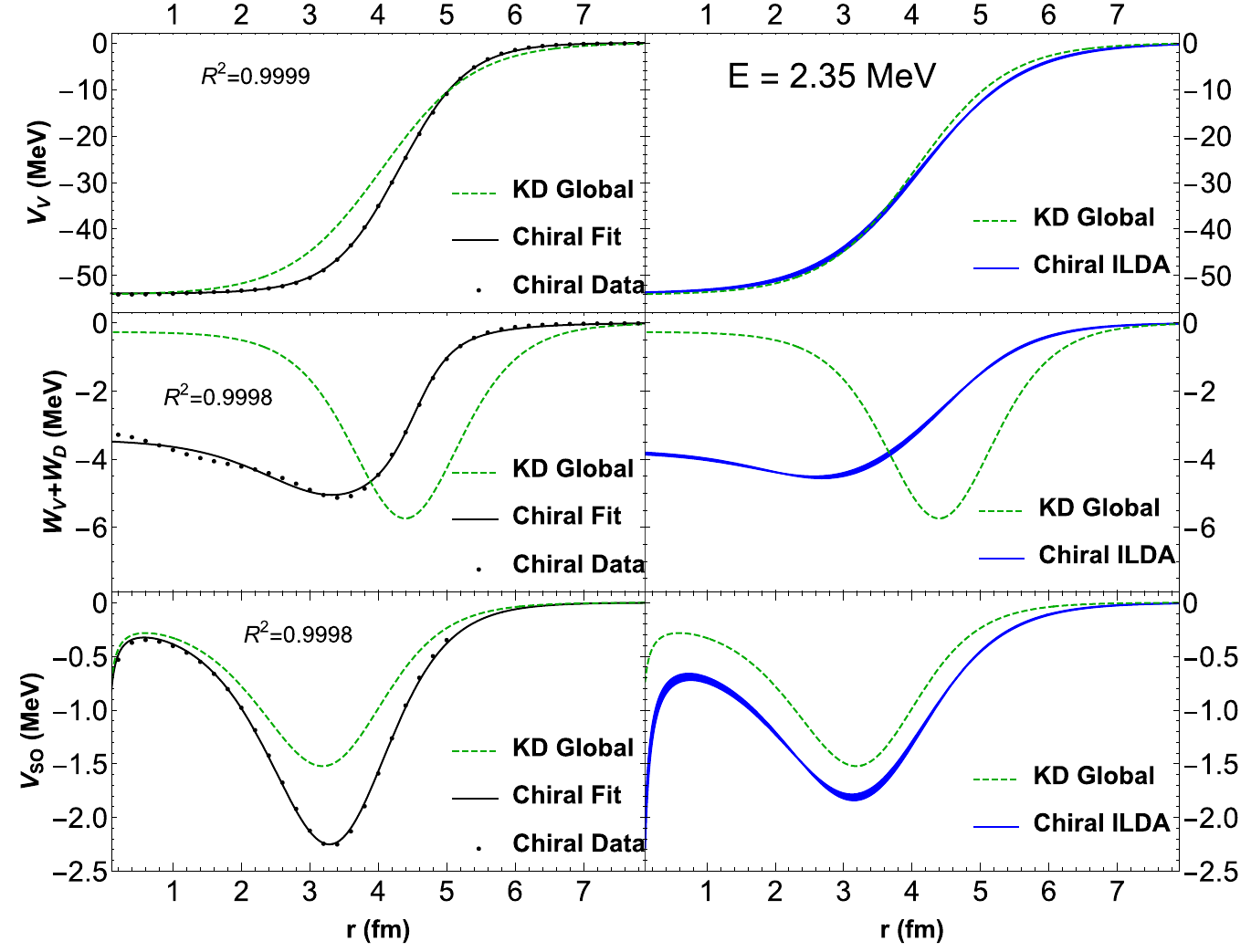}
\caption{Radial dependence of the real, imaginary, and spin-orbit contributions to the proton-$^{40}$Ca optical potential at projectile energy 2.35 MeV. The green-dashed lines are from the phenomenological Koning-Delaroche \cite{koning03} optical potential. The dotted lines are from homogeneous nuclear matter calculations of the nucleon self energy from chiral effective field theory two-body and three-body forces \cite{machleidt11,coraggio14} combined with the local density approximation (left) and improved local density approximation (right). The solid lines are fits to the chiral effective field theory results using the Koning-Delaroche global parameterization. Reprinted figure with permission from Ref.\ \cite{whitehead19}, copyright (2019) by the
American Physical Society.}
\label{fig:ilda}
\end{figure}

Another problematic feature of the nuclear matter approach is that in the local density approximation, Eq.\ \eqref{omp}, implies that a projectile nucleon only begins to interact with the target nucleus once the two are in physical contact. Since the range of the nuclear force is approximately 2\,fm, the surface diffuseness in the local density approximation is too small. This can be improved by accounting for the finite range of the nuclear force using the improved local density approximation (ILDA):
\begin{equation}
{V}(E;r)_{ILDA}=\frac{1}{(t\sqrt{\pi})^3}\int V(E;r') e^{\frac{-|\vec{r}-\vec{r}'|^2}{t^2}} d^3r',
\label{eq:ilda}
\end{equation}
where a smearing function with adjustable length scale $t$ is introduced and taken to be the typical range of the nuclear force. In Fig.\ \ref{fig:ilda} is shown the difference between the real, imaginary, and spin-orbit contributions to the proton-$^{40}$Ca optical potential at projectile energy 2.35 MeV in the local density approximation (left) and improved local density approximation (right). The dotted lines are the results obtained from chiral effective field theory calculations of the nucleon self energy in homogeneous nuclear matter folded with the density distribution of $^{40}$Ca using a mean field model fitted to the equation of state from chiral effective field theory \cite{whitehead19}. The solid lines show the fits to the chiral effective field theory calculations using the parameterized form of the Koning-Delaroche (KD) global optical potential \cite{koning03}, and the dashed-green lines are calculated from the KD global optical potential. First, one notices that the phenomenological parametrization accurately fits the microscopic calculations. Second, it is evident that the main effect of the improved local density approximation is to increase the diffuseness of the optical potential surface. Finally, one sees that the microscopic imaginary part is much less surface-peaked than phenomenology would suggest, while at higher projectile energies the imaginary part becomes nearly a factor of two larger than the phenomenological depth \cite{whitehead19}.

\begin{figure}[t]
\begin{center}
\includegraphics[width=\textwidth]{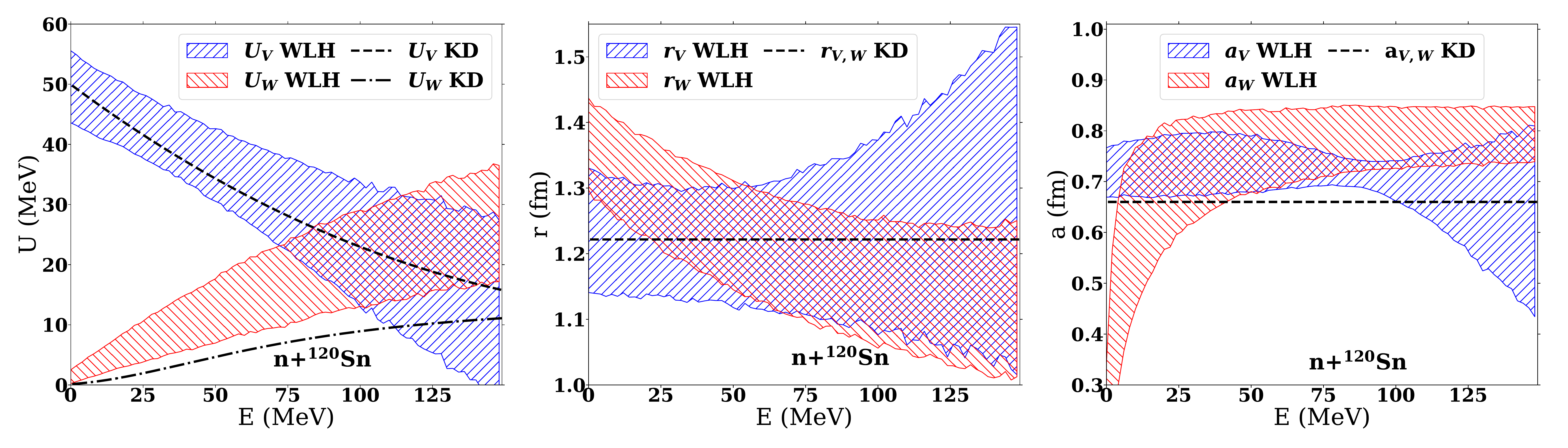}
\caption{ The depth (left), radius (middle), and diffuseness (right) parameters of the central real and imaginary terms in the WLH and KD global optical potentials shown as a function of energy. The bands of the WLH parameter distributions represent the 95\% credible interval from 5000 random parameter samples.
\label{fig_param}}
\end{center}
\end{figure}

Systematic uncertainties in the construction of microscopic global optical potentials based on the nuclear matter Green's function approach have been studied in Ref.\ \cite{whitehead21}. In this work, a set of five chiral N2LO and N3LO potentials with varying momentum-space cutoff values was used to construct a global optical potential, referred to as WLH, with associated theory uncertainties. From a covariance matrix analysis, the optical potential parameters are expressed in terms of a multivariate normal distribution. This distribution can then be sampled many times to produce uncertainty estimates for optical potential parameters and derived scattering observables. In Fig. \ref{fig_param} the 95\% probability contours for the WLH depth, radius, and diffuseness parameter distributions for the central real and imaginary terms are shown as a function of projectile energy. In general the predicted parameter values of WLH are close to KD, however, one finds large uncertainties inherent to microscopically derived optical potentials. The real depth of WLH is quite similar to KD while the imaginary depth from the nuclear matter approach is roughly a factor of two larger than that of KD. On average, the geometry parameters of WLH are also quite similar to those from KD, though in the microscopic approach there
\clearpage
\begin{figure}[t]
\begin{center}
\includegraphics[width=\textwidth]{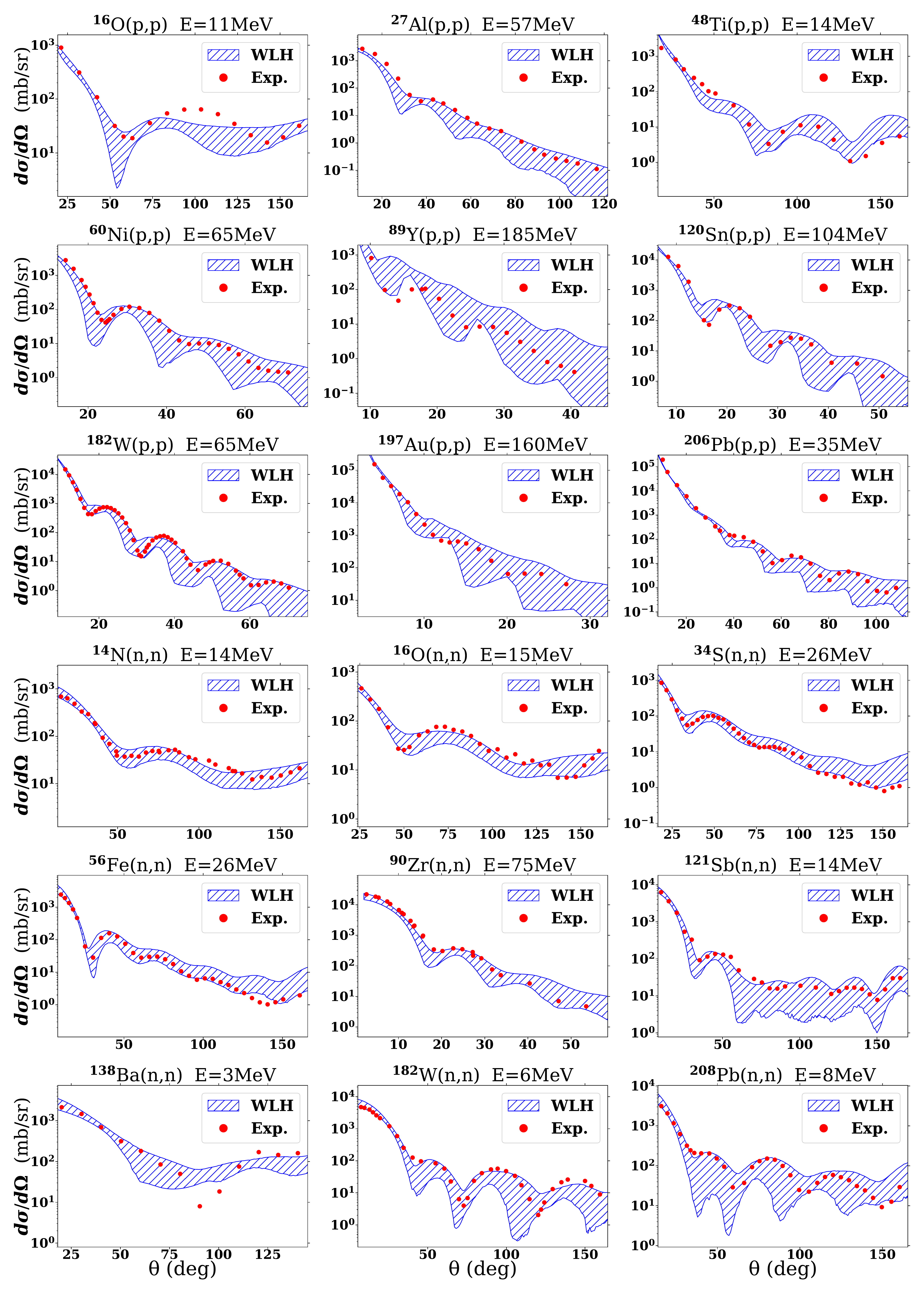}
\caption{ Proton and neutron elastic scattering cross sections predicted by the WLH global optical potential compared to data. The bands of the WLH results represent the 95\% credible interval of 5000 random parameter samples.
\label{fig_cs} }
\end{center}
\end{figure}
\clearpage
\noindent can be significant differences between the geometry parameters for the real and imaginary parts, in contrast to the KD parameterization which assumes the real and imaginary volume geometry parameters are identical.

\begin{figure}[t]
\begin{center}
\includegraphics[scale=0.234]{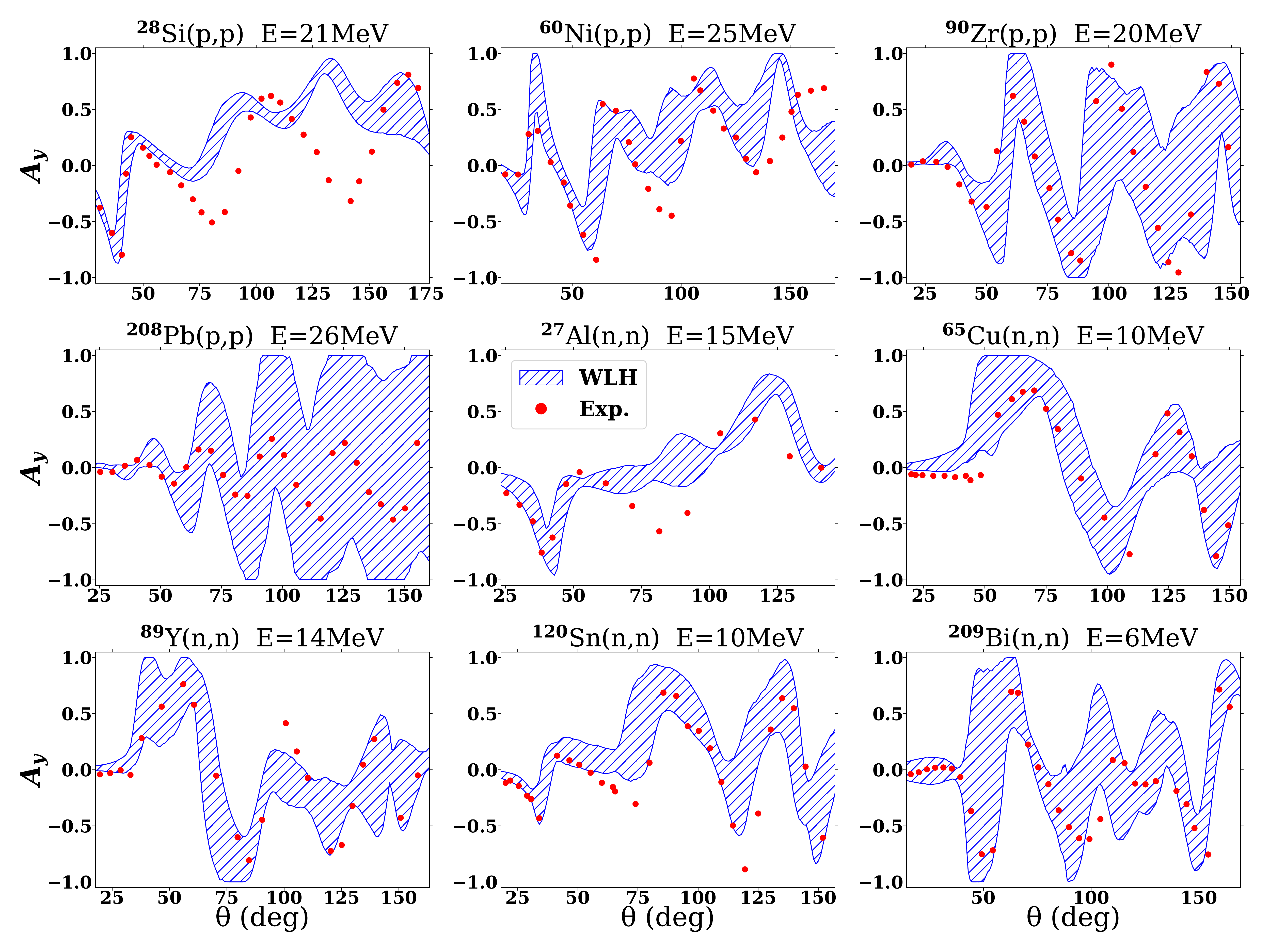}
\caption{ Proton and neutron analyzing powers predicted by the WLH global optical potential compared to data. The bands of the WLH results represent the 95\% credible interval of 5000 random parameter samples.}
\label{ap}
\end{center}
\end{figure}

In Fig. \ref{fig_cs} elastic differential cross sections for both proton and neutron projectiles predicted by WLH are compared to experimental data for a wide range of targets and energies. In general the WLH optical potential reproduces elastic scattering well for projectile energies of $E \lesssim 150$ MeV. As the projectile energy increases to $E \sim 200$ MeV, the predictions of WLH tend to have more discrepancies with data and larger uncertainties. In Fig. \ref{ap}, analyzing powers from the WLH microscopic global optical potential for proton and neutron projectiles are shown for a range of energies and target isotopes. Again, the overall agreement with experimental appears satisfactory within the theory uncertainties.

\section{\textit{Dispersive optical potentials}}

A great deal of experimental data exists to probe the nuclear mean field (i.e., optical potential) at a continuum of positive energies. At negative energies the nuclear mean field (i.e., shell model potential) affects the single-particle level structure of nuclei and the bound-state eigenfunctions, all at discrete values of the energy. Historically, the development of optical potentials and shell model potentials evolved independently. The aim of the so-called dispersive optical model \cite{mahaux86} is to provide a phenomenological framework to treat elastic scattering, reaction, and structure data all on an equal footing connected by a nuclear mean field that is continuous from negative to positive energies. The starting point is the dispersion relation linking the real and imaginary parts of the optical potential through the requirement of causality, where the scattering wave cannot be emitted before the arrival of the incident wave. This gives rise to Kramers-Kronig relations between the real and imaginary parts of the mean field:
\begin{align}
\label{dispV}
V(k;E) &= \frac{1}{\pi}{\cal P}\int_{-\infty}^\infty \frac{W(k,E^\prime)}{E^\prime - E}dE^\prime \\
\label{dispW}
W(k;E) &= -\frac{1}{\pi}{\cal P}\int_{-\infty}^\infty \frac{V(k,E^\prime)}{E^\prime - E}dE^\prime,
\end{align}
where ${\cal P}$ denotes the principal value integral. Thus, one sees that the real nuclear mean field in one energy regime is governed by the absorptive imaginary part across both regimes. Purely phenomenological optical potentials fit separately the real and imaginary components without respecting their mutual constraints through dispersion integrals. Such optical potentials therefore fundamentally violate causality.

In the dispersive optical model, one starts by breaking up the real part of the self energy into two parts
\begin{equation}
V(r,E) = V_{HF}(r,E) + \Delta V(r;E).
\label{domHFD}
\end{equation}
In principle, the energy dependence of the Hartree-Fock contribution $V_{HF}(r,E)$ arises by replacing the initially energy-independent and nonlocal part of the optical potential by its energy-dependent local equivalent \cite{perey62}. In practice, however, the Hartree-Fock contribution in the dispersive optical potential is often taken to be of the form in Eq.\ \eqref{domHFD} from the start. The energy-dependent (or ``dispersive'') part of the optical potential $\Delta V(r;E)$ that arises from temporal nonlocality is related to the imaginary part of the self energy through either Eq.\ \eqref{dispV} or the subtracted dispersion relation
\begin{equation}
\Delta V(r;E) = \frac{1}{\pi}{\cal P}\int W(r,E^\prime) 
\left ( \frac{1}{E^\prime - E} - \frac{1}{E^\prime - E_F} \right ) dE^\prime,
\end{equation}
which has several advantages, such as minimizing the magnitude of the dispersive correction in the vicinity of the Fermi surface, $\Delta V(r;E_F) = 0$, and taming the effects from large variations in the imaginary part of the optical potential near the Fermi surface.

Since only the real dispersive part $\Delta V(r,E)$ of the optical potential is fixed from the dispersion relation, a parameterized model for both the real Hartree-Fock $V_{HF}(r,E)$ contribution and the imaginary $W(r,E)$ contribution is needed. There is considerable flexibility when choosing the functional dependence on the energy $E$ and on the nuclear radius $R$ and diffuseness $a$ parameters. The following discussion will focus only on those features that are common to many of the different analyses. The Hartree-Fock term is typically written in the form
\begin{equation}
V(r,E) = V_{HF}^{vol}(E) f(r,r^{HF},a^{HF}) + V_{HF}^{sur} \frac{d}{dr}f(r,r^{HF},a^{HF}) + V_C(r) + V_{so}(r,E),
\label{domv}
\end{equation}
consisting of volume, Coulomb, and spin-orbit contributions as well as an optional surface real part. Most purely phenomenological optical potentials do not include a surface real term, which in recent dispersive optical model analyses \cite{dickhoff18} was found to be important to describe high-energy elastic scattering. The imaginary component is parameterized as
\begin{equation}
W(r,E) = W^{vol}(E) f(r,r^{vol},a^{vol}) + 4a^{sur}W^{sur}(E) \frac{d}{dr}f(r,r^{sur},a^{sur}) + W_{so}(r,E),
\label{domw}
\end{equation}
consisting of the usual volume, surface, and spin-orbit terms. The presence of an imaginary spin-orbit contribution also implies a dispersive correction $\Delta V^{so}(r,E)$ to the real spin-orbit interaction $V_{so}(r,E)$ in Eq.\ \eqref{domv}.

While the dispersive optical model is primarily phenomenological, some of the assumptions that go into defining the parametric form of the potential are motivated by microscopic calculations. For instance, general arguments indicate that the imaginary part of the self energy is symmetric about the Fermi energy and in fact vanishes at $E=E_F$. This property is built into the energy dependent parameters in Eq.\ \eqref{domw}. Far away from the Fermi energy, however, the imaginary part of the dispersive optical potential is typically allowed to break symmetry about $E_F$. In addition, microscopic calculations of the Hartree-Fock contribution find that the real volume component is a very smooth function from positive to negative energies, and hence one typically takes a low-order polynomial in powers of $E-E_F$.

\begin{figure}[t]
\begin{center}
\includegraphics[scale=0.81]{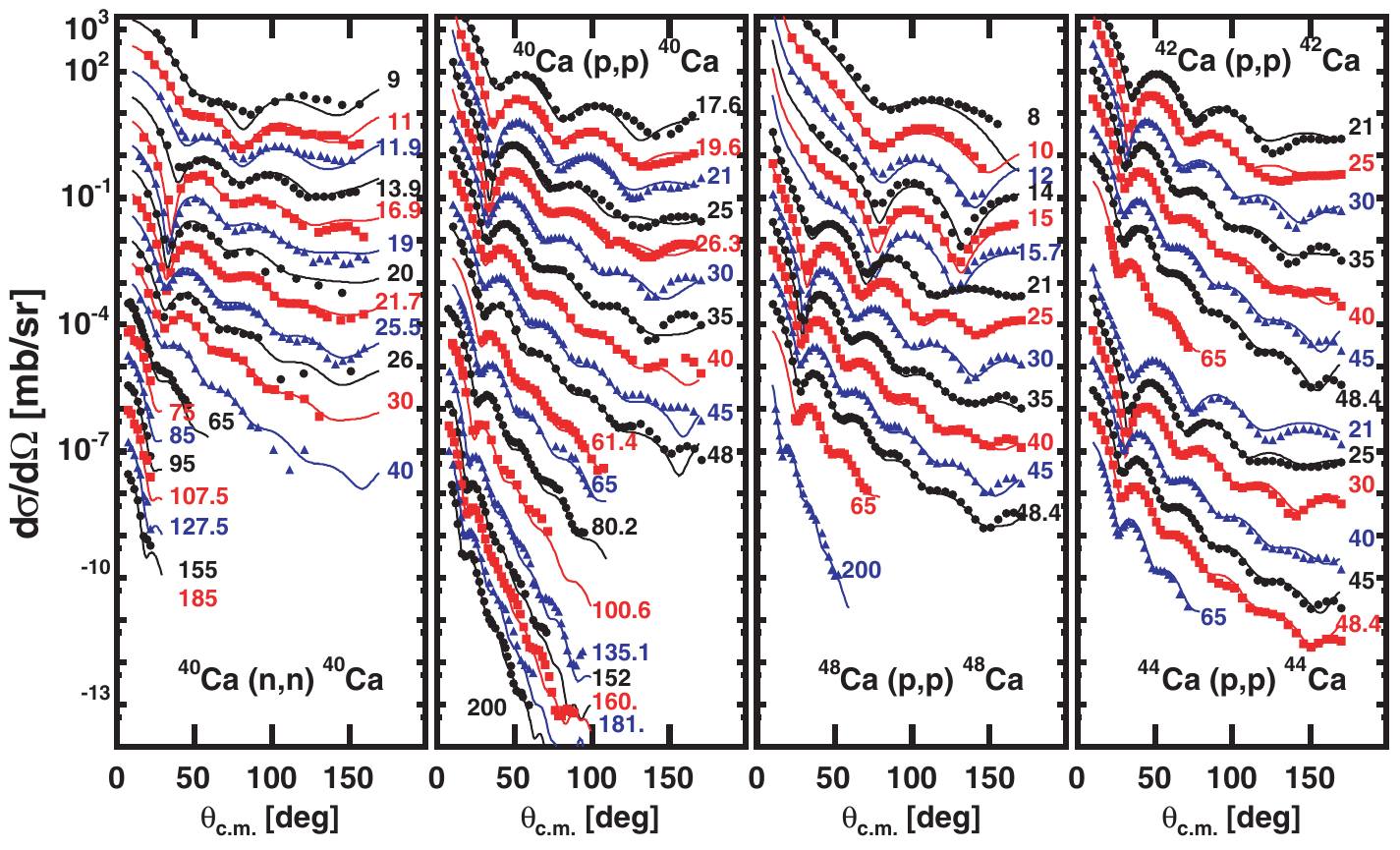}
\caption{Fitted differential elastic scattering cross sections for proton and neutron projectiles at varying energies on calcium isotopes within the dispersive optical model framework. Reprinted figure with permission from Ref.\ \cite{charity07}, copyright (2007) by the
American Physical Society.}
\label{fig_dom}
\end{center}
\end{figure}

In practice, the parameters that define the energy dependence and shape parameters of the real Hartree-Fock and imaginary components of the dispersive optical potential can be fitted to a wide range of scattering and bound-state data. For instance, single-particle energy levels, experimental one-particle removal widths, binding energies, charge densities, charge radii, differential elastic scattering cross sections, analyzing powers, reaction cross sections, and total cross sections \cite{charity07,pruitt20} have all been used in fitting the parameters of modern dispersive optical potentials. In Fig.\ \ref{fig_dom} is shown the results \cite{charity07} of fitted nucleon-nucleus elastic scattering cross sections for proton and neutron projectiles at varying energy on calcium isotopes within the dispersive optical model. One finds that it is possible to describe a large amount of elastic scattering data within this phenomenological framework, and similar quality of fits can also be obtained simultaneously for analyzing powers and reaction cross sections.

In this review, some of the modern theoretical tools for constructing nucleon-nucleus optical model potentials for nucleon elastic scattering and reactions have been outlined. Ab initio methods, such as self-consistent Green's function theory and the coupled-cluster method, can at present be applied at low projectile energies and for doubly-magic nuclei. In order to describe a wider range of target nuclei and projectile energies, one must make simplifying assumptions as in the nuclear matter Green's function approach or multiple scattering theory. These methods allow for the implementation of realistic nuclear two-body and many-body forces, which nowadays is firmly grounded in the low-energy realization of QCD, chiral effective field theory. Ab initio and microscopic reaction theory models can be linked to the large corpus of nuclear structure studies using the same underlying nuclear force models. A phenomenological framework for consistently analyzing both positive-energy scattering states and negative-energy bound states is provided by the dispersive optical model, which in addition respects microscopic constraints such as causal dispersion relations and the functional form of the imaginary optical potential near the Fermi surface.



\section{\textit{Acknowledgments}}
The authors thank M.\ Vorabbi for useful discussions and providing Figure \ref{fig_12C_sigma}. The authors also thank C.\ Elster and R.\ C.\ Johnson for helpful feedback on our manuscript. JWH is supported by the National Science Foundation under Grant No.\ PHY1652199 and by the U.S.\ Department of Energy National Nuclear Security Administration under Grant No.\ DE-NA0003841. TRW is supported in part by the U.S.\ Department of Energy (Office of Science, Nuclear Physics) under grant DE-SC0021422.


\end{document}